\documentclass[]{spie}  

\usepackage{amsmath,amsfonts,amssymb}
\usepackage{graphicx}
\usepackage[colorlinks=true, allcolors=blue]{hyperref}

\usepackage[perpage]{footmisc}

\title{The eXtreme Wavefront Control Toolkit: High-Contrast Imaging Instrument Control for Ground and Space-Based Coronagraphs }

\author[a]{Jared R. Males}
\author[b]{Joseph D. Long}
\author[c]{Sebastiaan Y. Haffert}
\author[a]{Kyle Van Gorkom}
\author[a]{Parker Johnson}
\author[c]{Rico Landman}
\author[a,d]{Eden McEwen}
\author[a,d,e,f]{Olivier Guyon}
\author[e,g]{Vincent Deo}
\author[a,e]{Miles Lucas}
\author[a]{Irina Stefan}
\author[a,d]{Katie Twitchell}
\author[a,d]{Jay Kueny}
\author[a,d]{Joshua Liberman}
\author[c]{Adam K. Taras}
\author[a]{Adam Schilperoort}
\author[c]{Matthijs Mars}
\author[a]{Ewan S. Douglas}
\author[a]{Laird M. Close}

\affil[a]{Steward Observatory, University of Arizona}
\affil[b]{Flatiron Institute, New York}
\affil[c]{Leiden Observatory, Leiden University, The Netherlands}
\affil[d]{James C. Wyant College of Optical Sciences, University of Arizona, USA}
\affil[e]{Subaru Telescope, National Astronomical Observatory of Japan}
\affil[f]{Astrobiology Center, National Institutes of Natural Sciences, Japan}

\affil[g]{Optical Sharpeners SAS, France}

\authorinfo{Further author information: (Send correspondence to J.R.M.)\\J.R.M.: E-mail: jrmales@arizona.edu}
\begin{document} 
\maketitle

\begin{abstract}
We present the eXtreme Wavefront Control Toolkit (XWCTk), a comprehensive instrument control software system developed for the MagAO-X extreme adaptive optics (ExAO) instrument. The XWCTk is built on a foundation of the ImageStreamIO (ISIO) / MILK / CACAO low-latency image processing and high dimensional control tool chain. Higher level instrument control is managed with the Instrument Neutral Distributed Interface (INDI). These tools are combined via a modern c++ application framework which seamlessly provides configuration, logging, distributed inter-process communication (IPC) with INDI and low-latency IPC with ImageStreamIO. The core library provides various components, e.g. standard hardware interfaces, which can be added to an application as needed. On the MagAO-X instrument, every detector (including in science focal planes) is a potential wavefront sensor capable of sending commands to three separate deformable mirrors. MagAO-X utilizes a distributed control system, where a separate real-time controller (RTC) and the instrument control computer (ICC) each manage low-latency wavefront control tasks (up to 3.6 kHz on the 2000 actuator main AO system, and over 8 kHz on the low-order coronagraph system) but are capable of coordinated control.  Implemented algorithms include neural networks for nonlinear reconstruction of wavefronts at over 3 kHz. We have recently begun incorporating distributed raspberry pis for accelerometer data acquisition at various locations on the telescope, with low-latency streaming to the real-time computers for sensor fusion control. A core design principle of the XWCTk is that all data can be saved all the time. This includes full-rate WFS images, deformable mirror commands, as well as science data. To facilitate this we have implemented a custom lossless compression system capable of sustaining high data rates to disk. A python interface for scripting and experimentation, as well as a python application framework is provided which can be used for non-real-time tasks. Remote operations (e.g. from Tucson Arizona when the instrument is at LCO in Chile) is routine. The XWCTk is under continuous development for the MagAO-X instrument, and will be adapted for GMagAO-X, the planned first-light ExAO coronagraph for the Giant Magellan Telescope. Additionally, XWCTk is the baseline for a space high contrast imaging instrument, and as such ongoing development is focused on automation for robust operation in flight.
\end{abstract}


\section{INTRODUCTION}
\label{sec:intro}  

Extreme Adaptive Optics (ExAO)\cite{2018ARA&A..56..315G} systems are highly complex, requiring control of thousands of actuators on a deformable mirror (DM) at speeds over 1 kHz.  This requires low-latency image processing at the same speed for wavefront reconstruction.  When combined with a coronagraph, such systems have many other degrees of freedom such as filter wheels and selectable coronagraph optics.  Coronagraphs typically have internal wavefront sensing and control (WFS\&C) systems operating at similar speeds\cite{2014PASP..126..586S}. Furthermore, the now routine use of focal plane WFS\&C (i.e. ``dark holes'') means that even the science detectors become part of precision WFS\&C loops\cite{2024SPIE13097E..09M}.  

With these characteristics, the software of an ExAO instrument harbors the most complexity in the system.  It is the software that must manage the many degrees of freedom, enable the high-performance image processing and support all phases of operations (from control system calibration to post-processing).  Added to this, ExAO systems tend to be experimental, undergoing frequent modifications and upgrades.  New algorithms are constantly in development and new higher-performance hardware frequently becomes available.  This means that the software system is also never finished.

Here we present the ``eXtreme Wavefront Control Toolkit'' (XWCTk), the software system developed to control the MagAO-X ExAO instrument on the Magellan Clay 6.5 m telescope at Las Campanas Observatory\cite{2024SPIE13097E..09M}.  MagAO-X has three DMs, a pyramid wavefront sensor (PyWFS) that operates at speeds up to 3.6 kHz, a Lyot-style coronagraph with many options, low-order WFS fed by both the focal plane mask (FPM) and Lyot stop, and dual-channel visible-wavelength science cameras as well as supporting visitor instrumentation.  There are least 19 presentations in these proceedings which include the use of MagAO-X \cite{lucas_spie_2026, tonucci_spie_2026_1, close_spie_2026_1,twitchell_spie_2026, mars_spie_2026_1, mcewen_spie_2026_1, mcewen_spie_2026_2,
       mars_spie_2026_2, kueny_spie_2026_1, landman_spie_2026_1, liberman_spie_2026, kueny_spie_2026, 
       long_spie_2026, haffert_spie_2026_1, gonzalez_spie_2026, patel_spie_2026, tonucci_spie_2026_2,
johnson_spie_2026, hom_spie_2026}.

Here we describe the XWCTk architecture, include the foundational technologies on which it is based, the core application framework, and the supporting libraries developed along with it.  We then describe the rich scripting interface that has been developed to support experimentation. We also describe the various user interfaces developed to support both operators and observers.  Finally, we present several example use cases and discuss the future of the XWCTk.

\noindent\textbf{Note on nomenclature:} As of this writing the source code for XWCTk is still under the MagAO-X source repo, and named for MagAO-X.  E.g., what we refer to here as ``XWCApp'' is called ``MagAOXApp'' in the live source.  We are in the process of refactoring the core library and supporting utilities, which includes renaming to ``XWC*''.  Here we use ``XWC'' in line with this soon-to-be released naming convention and the planned usage going forward.

\section{Toolkit Architecture}
\subsection{Included Technologies}
The XWCTk is built upon several pre-existing systems.  The two main ones are INDI and CACAO.
\subsubsection{INDI}
For control of the instrument we adopted the Instrument Neutral Distributed Interface (INDI)\cite{indi_specs}\footnote{\url{https://www.clearskyinstitute.com/INDI/INDI.pdf}}, a software system developed specifically for control of astronomical instrumentation. INDI is an XML-based publish-subscribe protocol for inter-process communication that enables application-to-application coordination, scripting, and user interface implementation in a distributed system. The core component of an INDI system is the \textit{indiserver}, one of which runs on each machine in the system.  On each machine, a set of applications connects to the \textit{indiserver} as a ``driver''.  Multiple machines are networked together, with their respective \textit{indiserver}s communicating such that the total collection of ``drivers'' appears as a single system.  Scripts and UIs (graphical or terminal) connect to an INDI system as ``clients''.

We based our INDI library on the version maintained for the Large Binocular Telescope Interferometer (LBTI), which includes \textit{indiserver} and the command line interface (CLI) utilities \textit{getINDI} and \textit{setINDI}.  We also have adapted LBTI's \textit{libcommon} c++ library for developing INDI applications, making a stripped-down version that focuses only on INDI protocol implementation.

\subsubsection{CACAO}
The adaptive optics control system is based on the Compute and Control for Adaptive Optics (CACAO)\cite{2018SPIE10703E..1EG,deo_spie_2026} framework.  CACAO itself consists of three components:

\noindent\textbf{ImageStreamIO:} The foundation of the CACAO system is the ImagesStreamIO (ISIO) shared memory format and supporting functionality\footnote{\url{https://github.com/milk-org/ImageStreamIO}}.

\noindent\textbf{MILK:} On top of ISIO sits the Multi-purpose Imaging Libraries toolKit (MILK)\footnote{\url{https://github.com/milk-org/milk}}.  MILK provides a multi-process image-processing framework for using ISIO image-streams.  This includes process management tools and a configuration system.

\noindent\textbf{CACAO:} CACAO itself\footnote{\url{https://github.com/cacao-org/cacao}} is installed as a plugin to the MILK system.  CACAO provides high-performance low-latency compute modules for AO control.  This includes calibration acquisition, system characterization tools (e.g. latency measurement), and modal control.  

\subsection{Application Framework}
\begin{figure}[h!]
    \centering
    \includegraphics[width=3in]{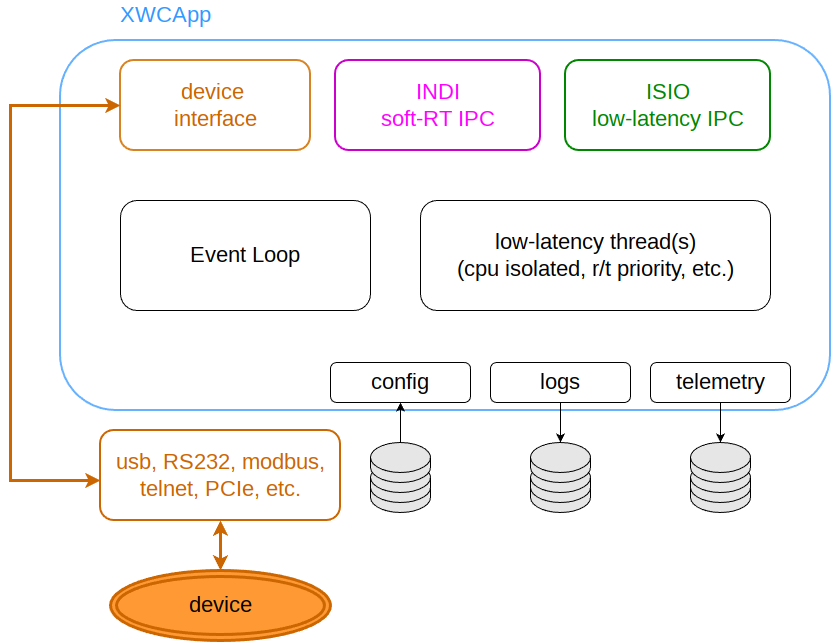}
    \caption{The basic component of the XWCTk is the ``XWCApp''.  Each instance of an XWCApp has standard configuration, logging, and telemetry.  INDI communications is optional at compile time but is usually included.  ImageStreamIO (ISIO) low-latency IPC is available through several selectable interfaces depending on the purpose of the app.  The standard event loop is specialized for the purposes of the app (see Fig. \ref{fig:xwcapp_fsm}).  Finally, a library of device control modules are available implementing interfaces to common components.  \label{fig:xwcapp}}
\end{figure}

XWCTk is organized around an application framework, the base of which is the c++ class \textit{XWCApp}.  All application classes are then derived from this base.  The base class implements many standard features, including PID locking, privilege management, configuration, logging, IPC with INDI, and thread management including real-time priority and cpuset isolation.  See figure Figure \ref{fig:xwcapp}. Additional features can be added by including other base classes, such as telemetry and low-latency IPC with ISIO.  These additional features are included using the curiously recursive template pattern (CRTP) idiom of c++ which avoids the diamond problem of multiple inheritance.  Thus one application can be connected to many ISIO image streams, for instance.

\begin{figure}[h!]
    \centering
    \includegraphics[width=5in]{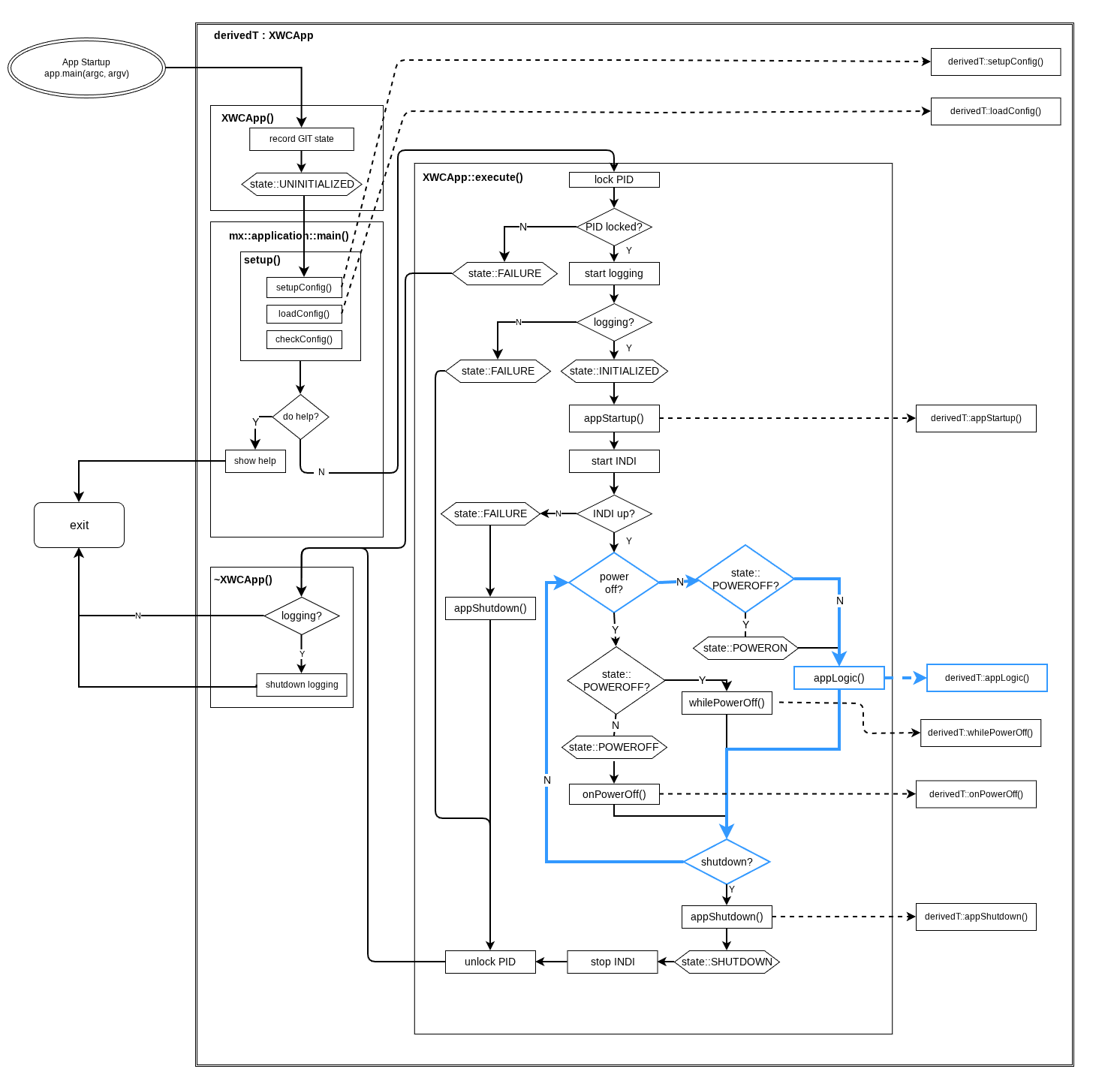}
    \caption{The standard finite state machine (FSM) logic of an XWCApp.  The derived application implements the functions shown at right, connected by dashed lines, which are called from within the core FSM.  The main event loop is highlighted in blue.   \label{fig:xwcapp_fsm}}
\end{figure}

The \textit{XWCApp} implements a standard execution path and event loop.  The standard execution function coordinates startup and starts the event loop.  The loop keeps application state up to date, monitors power state (through INDI subscription to the power controllers) and calls the derived class's \textit{appLogic} function once per second.  The loop responds to shutdown requests and then coordinates shutdown.  The specialized logic for each app is implemented in the derived classes functions, notably: \textit{appStartup}, \textit{appLogic}, and \textit{appShutdown}. See Figure \ref{fig:xwcapp_fsm}.

\begin{figure}[h!]
    \centering
    \includegraphics[width=6.5in]{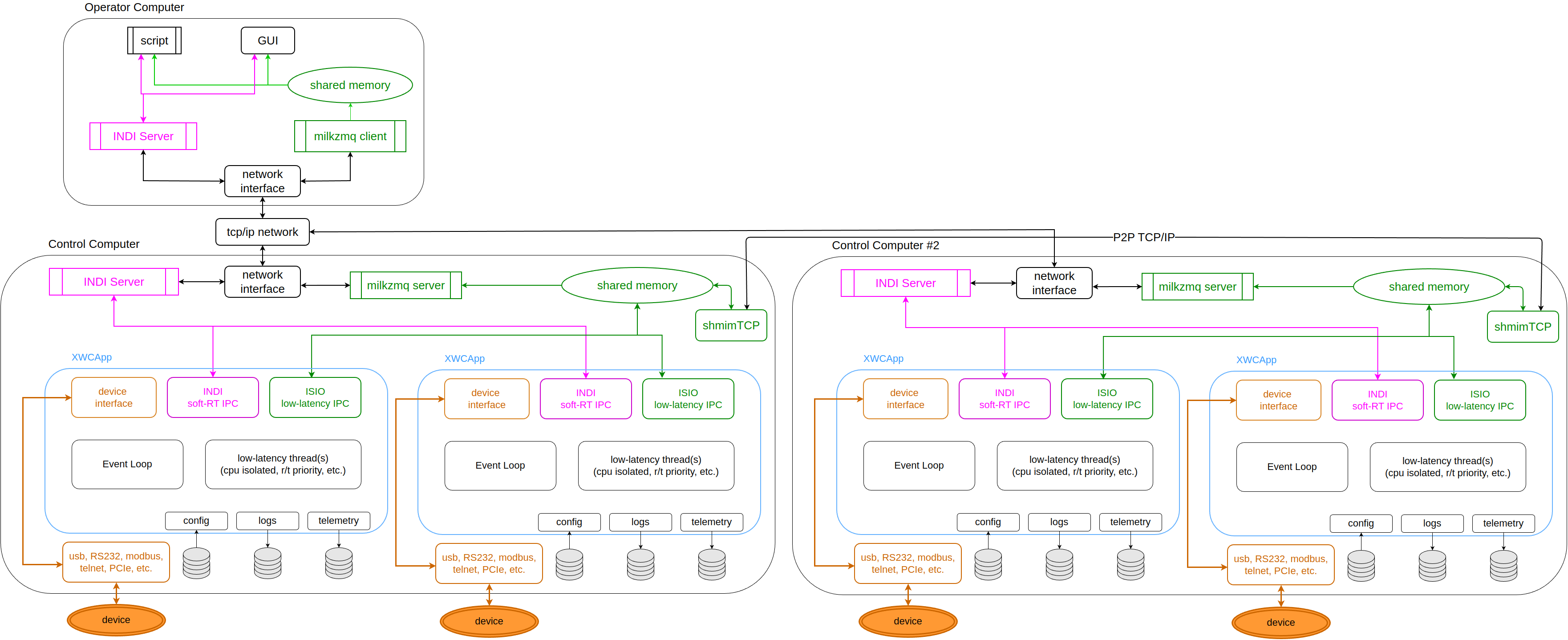}
    \caption{Building a distributed control system out of XWCApps.  The INDI protocol and tools for synchronizing and transmiting ISIO shared memory streams enable multiple hardware-controlling computers to appear as a single instrument.  \label{fig:xwcapps_connections}}
\end{figure}

A typical XWCTk-based control system is distributed across multiple computers.  Each XWCApp is written to perform a single task, such as ``control a filter wheel'' or ``calculate an image centroid''.  As shown in Figure \ref{fig:xwcapps_connections} the apps on a given computer control and interact with the hardware connected to that computer.  The apps on different computers are connected via INDI for normal command and control.  Image data is shared across computers using the ``shmimTCP'' utility from CACAO, which synchronizes ISIO streams across the network (typically a dedicated point-to-point connection).  Operator work stations are connected using INDI.  

\subsubsection{Standard Interfaces}

The framework includes a collection of standardized interfaces which can be inherited as base classes through the CRTP.  Doing so requires that the XWCApp instance provide a specific interface (set of members and functions) to the base class, and that the parent class call the base class member functions (e.g. \textit{bass::appLogic} is called within the parent's \textit{appLogic}).  As an example, all camera control apps inherit from the \textit{stdCamera}.  This then ensures that all cameras have the same configuration conventions and expose the same INDI properties.  This in turn makes implementation of flexible algorithms, scripts, and UIs very straightforward.

\subsection{Co-Developed Libraries}
Several supporting libraries have been developed to support the XWCTk.

\subsubsection{flatlogs}
For process logging and telemetry we adopted the flatbuffers\footnote{\url{https://github.com/google/flatbuffers}} binary serialization protocol due to its memory efficiency.  The \textit{flatlogs} library uses flatbuffers for the payload of a log, to which is prepended a header containing a message type identifier, priority, and timestamp.  The \textit{flatlogs} library provides c++ helpers for code generation and encoding, decoding, and formatting of log and telemetry messages.  Related tooling includes the ``logdump'' utility which prints logs from a given process.  This is aliased as ``teldump'' for viewing the identically formatted telemetry files.

\subsubsection{Compression with xrif}
\begin{figure}
    \centering
    \includegraphics[width=4in]{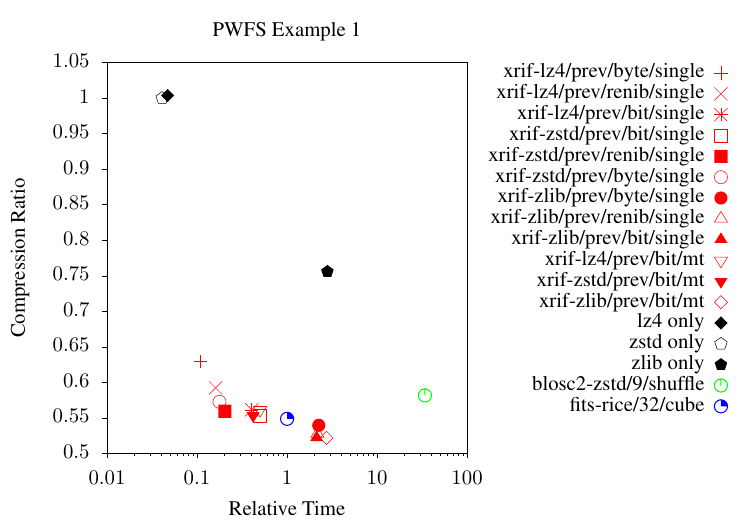}
    \caption{Comparison of the xrif compression system (red) to other codecs.  The xrif points correspond to various configurations of bit-shuffling (whole bytes, 4-bit nibbles, or bit-by-bit) and final codecs (zlib, lz4, or zstd), and single threaded vs multi-threaded (mt).  The blue-circle with notch corresponds to the Rice algorithm used in cfitsio.  \label{fig:caption}}
\end{figure}

The MagAO-X project adopted a requirement to save all data from all sources all the time during on-sky operations.  The rationale for this was to support telemetry-based point spread function (PSF) estimation for high-contrast post-processing\cite{long_spie_2026}.  In order to minimize disk usage and ensure that this requirement could be met we developed the eXtremely Reduced Image Format (xrif\footnote{\url{https://github.com/jaredmales/xrif}}), a lossless compression system for integer data.  It combines delta encoding (differencing either image to image or pixel to pixel), bit-shuffling (re-ordering post-differencing bits by their significant), and a standard codec such as zlib, lz4, or zstd.  After differencing and shuffling, these codecs produce excellent compression ratios.  Figure \ref{fig:caption} compares the performance of various configurations of xrif to other codecs when compressing MagAO-X PyWFS data.  xrif delivers compression as good or better than the Rice algorithm used in cfitsio.  When optimized for speed using the lz4 codec, xrif compression is somewhat worse but 10 times faster.

\section{Instrument scripting API}

As a testbed for novel software and hardware approaches to high-contrast imaging, we need a lightweight interface to enable experimentation. This is implemented with two pieces of our software: PurePyINDI and the magaox-python package. Those in turn underlay a web-based interface for guest observers that gives an overview of the instrument's status at a glance and lets observers name and save their data.

\subsection{PurePyINDI}

PurePyINDI footnote{\url{https://github.com/xwcl/purepyindi2}} is a from-scratch Python implementation of the INDI protocol. A typical experiment through the scripting interface involves a Jupyter notebook, served through a JupyterLab interface on the relevant machine within MagAO-X. From that interface, the experimenter can use PurePyINDI to connect a client and interactively set and query properties of the instrument.

With version 2.0 of PurePyINDI, we added the ability to implement ``devices'' as Python programs. These device processes, or ``drivers'' in the language of the INDI standard, use an event loop structure similar to other XWTK applications and callbacks to handle commands and property updates. We have used the Python scripting interface to develop active atmospheric dispersion control\cite{twitchell_spie_2026}, as well as the audible alerts subsystem to improve operator alertness. In both cases, the ability to reuse the extensive Python package ecosystem accelerated development.

\subsection{magaox-python}

To implement capabilities that are more closely tied to the XWTK architecture, we have a Python package in the main MagAO-X git repository called (fittingly) ``magaox''. This uses PurePyINDI under the hood for commanding, as well as ImageStreamIOWrap from CACAO to enable reading and writing shmims to capture images or control deformable mirrors. The interface exposed by magaox-python is compatible with HCIPy\cite{2018SPIE10703E..42P} and indeed they are frequently used together for things like turbulence screen generation and image processing.

\section{User Interfaces}

The INDI protocol was designed to enable flexible user interfaces.  We have made use of this to build a suite of tools for operator control of an XWCTk system.

\subsection{Terminal Interfaces}

At the lowest level one can interact with the system with the \textit{getINDI} and \textit{setINDI} CLI.  We also developed an \textit{ncurses} base utility called \textit{cursesINDI}. Using the introspective capabilities of INDI, \textit{cursesINDI} presents a whole-system view of all properties and through keyboard shortcuts allows interaction with each property.     

\subsection{Qt Widgets}

The XWCTk includes a Qt-based graphical user interface (GUI) library.  This melds the \textit{libcommon} INDI-client framework with a set of standard re-usable Qt widgets.  We harness the standard interfaces of XWCTk to build flexible reusable GUIs.  For example, the camera control GUI adapts itself to the exposed properties of a given camera, populating only those widgets needed for that camera.  The result is a standard interface presented to users without exposing unnecessary controls.

\begin{figure}[h!]
    \centering
    \includegraphics[width=3in]{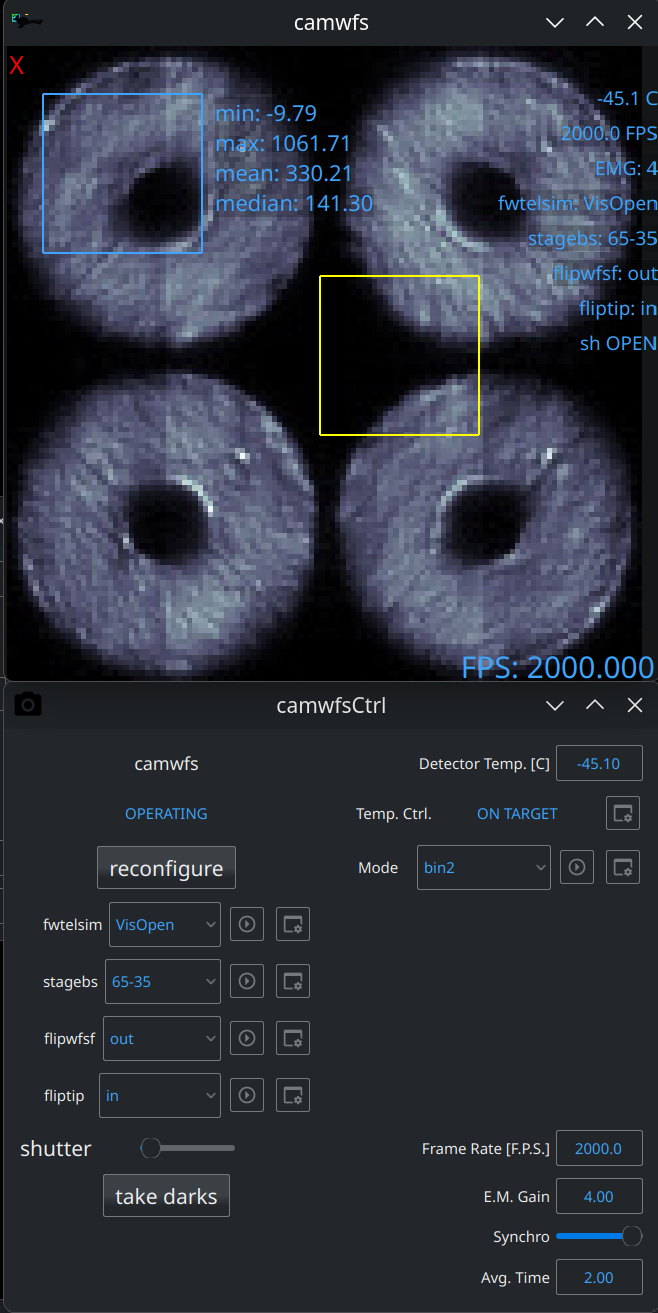}
    \caption{A typical XWCTk GUI, showing image display with \textit{rtimv} with overlays and the camera control GUI for the PyWFS of MagAO-X.  \label{fig:gui}}
\end{figure}

\subsection{rtimv}

The main image display tool is called \textit{rtimv}\footnote{\url{https://github.com/jaredmales/rtimv}}, which was originally developed for the MagAO project's VisAO camera.  See Figure \ref{fig:gui}. It is focused on the efficient display of high-frame rate camera images, with ds9-like tools for control and analysis\footnote{\url{https://jaredmales.github.io/rtimv-doc/md_UserGuide.html}}.  On a local computer, \textit{rtimv} connects directly to an ISIO shared-memory image and updates the display when the image is updated, though usually decimated to avoid unnecessary use of processing (i.e. we only show a 2 kHz WFS camera at 30 FPS). It also displays FITS files.

\textit{rtimv} also has a client-server mode.  Here image rendering is done on the server, which is normally an instrument computer.  The resulting jpeg image, with compression ratios of 2:1 to over 30:1 depending on the source data and display configuration, is sent to the client.  Additionally, any information that requires true raw-pixel values, such as the brightness of a pixel or statistics in a region, is calculated on the server and sent to the client.  The interaction between client and server is managed using gRPC.  The client behaves identically to the local version, such that operation of the software is identical in remote operations (discussed below).

Using the Qt plugin system, we have developed a suite of plugins for \textit{rtimv}.  This includes an INDI client, and display overlays to show INDI-derived information such as camera configuration, loop status, and the status of data saving for that camera.  Such an overlay is visible along the upper right edge in Figure \ref{fig:gui}.

\subsection{sup Web Interface}

We have developed a web browser interface for maximum flexibility, called \textit{sup}\cite{2022SPIE12185E..3PL}. The sup interface is a Python application on the backend and a VueJS ``single-page-app'' frontend, running in the user's web browser. The backend serves primarily to proxy INDI commands between the user's browser and an instance of the PurePyINDI client connected to the instrument. With each device within the instrument responsible for tracking its own state and reporting when queried, we don't have to worry about representations getting ``out of sync'' between the web interface and the Qt-based GUIs.

Within this web interface, users can see the instrument's current state, but not every controllable property has a corresponding control. This is intentional, as the primary audience for this interface is guest observers. The guest observer is given the necessary control to name their datasets and track their target's altitude, as well as starting and stopping the saving of camera and deformable mirror data.

Since guest observers access the instrument over a VPN, they are able to load this page in their browser at home and verify that the instrument is configured as they expect without installing the full software suite.

\section{Multi-developer Hacking and AI}

The XWCTk has to-date primarily been developed by a few members of the MagAO-X team, but recently the number of scientists working on new features has grown significantly. This includes many junior members of the team such as PhD students and post-docs as well as professional software engineers.  For any instrument, there is ultimately only one set of hardware to control and test on, and the real test for any new algorithm is at the telescope on-sky, where we must maintain traceability, reproducibility, and high reliability for normal observations. This complicates the usual git workflow of feature branches for parallel development, because we must eventually run the code on the actual instrument.  To manage this complexity, we use a development policy\footnote{\url{https://magao-x.org/docs/handbook/software/hacking.html}} that requires only committed code be used on-sky and sets rules for where code is edited and updated on the instrument.

Relatedly, we have developed a policy for the use of artificial intelligence (AI) in the writing of code\footnote{\url{https://magao-x.org/docs/handbook/software/agents.html}}.  We encourage the use of these time-saving tools, but seek to maintain engineering rigor and documentation.  We do this through the documentation of plans and prompts as commmitted files in the repository, while adhering to the standard development processes.

\section{XWCTk Use}

\subsection{Remote Operations}

Since 2024 MagAO-X is normally set up in the cleanroom at LCO when not mounted on the telescope.  This is done to enable remote experiments and optimization of the XWCTk. Users connect from Tucson, AZ and from as far away as Leiden, The Netherlands.  Beginning in April, 2026, MagAO-X now has the capability to be installed on the telescope for months at a time, with the MagAO-X team supporting observations remotely from Tucson.  This has been enabled by several aspects of the XWCTk design.  Remote desktops are not used, rather we run the same GUIs on a local workstation and connect them to the \textit{indiserver} over an ssh tunnel.  The client-server mode of \textit{rtimv} is also used.  The result is that the user experience is extremely similar during remote operations, with only a small amount of extra latency that is seldom noticed.  With only very occasional brief slowdowns,  image display is very responsive.  We observe the full 30 FPS display rate of the WFS images in Tucson, with a measured one-way time in the few hundred ms.  As a result we have had little difficulty managing the highly dynamic operation of MagAO-X in closed-loop on-sky.

\subsection{Neural Network WFS Reconstruction}

The PyWFS is inherently nonlinear, and as such the linear reconstruction typically employed prevents achieving optimal sensitivity\cite{mcewen_spie_2026_1}. One consequence of this is the need to modulate the beam on the pyramid tip, which opto-mechanically places speed limits on the system due to limitations of hardware.  We have recently demonstrated the use of a convolutional neural network (CNN) on-sky on MagAO-X to reconstructed the unmodulated (fully non-linear) PyWFS\cite{2025A&A...696L...1L,landman_spie_2026_1}.  The architecture of XWCTk was exploited to seamlessly integrate the CNN reconstructor, implemented in an XWCApp with TensorRT in c++, into the control loop.  Once the CNN was running and producing modal coefficients, it was a simple matter of changing symbolic links to the ISIO shared memory images to bypass the linear reconstructor close the loop on the CNN outputs. No other changes to the hardware control or loop processes were required.

\subsection{Data Acquisition with Raspberry Pis}
\begin{figure}[h!]
    \centering
    \includegraphics[width=3in]{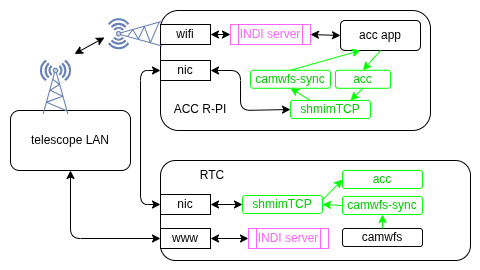}
    \caption{Architecture of the Raspberry Pi based accelerometer data acquisition system.  The XWCTk is installed on the RPi and used for low-latency data acquisition.  The measurements are synchronized with the WFS detector readout and transferred to the RTC for use in the AO control loop. \label{fig:accel}}
\end{figure}

We have recently extended the MagAO-X instrument to include accelerometers mounted on the Clay Telescope.  The goal of this project is to read accelerometers at high speed and use the outputs for real-time control of vibrations.  To this end we have developed a data acquisition system based on Raspberry Pi (RPi) single board computers, see Figure \ref{fig:accel}.  Using the Ubuntu Server operating system, we installed the XWCTk on the RPi and implemented an XWCApp to acquire readings from an analog to digital converter via the GPIO interface. Standard system components such as an ``indiserver'' are started on the RPi to integrate the accelerometer system into MagAO-X for command and control.  The accelerometer readings are published using ISIO for hard-R/T IPC, and the CACAO tool ``shmimTCP'' is used to transfer the readings over a dedicated point-to-point ethernet cable using the 1G interface on the RPi to the main MagAO-X RTC for inclusion in the control loop. The data acquisition is likewise synchronized with the WFS readout via the same point-to-point link also using ``shmimTCP''. See Johnson et al.\cite{johnson_spie_2026} in these proceedings for details about the performance and usage of this system.

\subsection{Ground-based Instruments}

In addition to MagAO-X, some parts of the XWCTk have been adapted for use on the MMT adaptive optics exoplanet characterization system (MAPS)\cite{montoya_spie_2026}.  The XWCTk is the baseline software system planned for use on GMagAO-X, the first-light ExAO system for the Giant Magellan Telescope\cite{males_spie_2026_1,close_spie_2026_2,haffert_spie_2026_2,giorgetti_spie_2026}.

\subsection{Testbeds}

The XWCTk is in use to control several high-contrast imaging testbeds.  These include the Comprehensive Adaptive Optics Coronagraph Testbed Instrument (CACTI)\cite{2022JATIS...8d9001S}, the Tiny Observatory for Telescope Optimization (TOTO)\cite{blomquist_spie_2026}, and the Space Coronagraph Optical Bench (SCoOB) testbed\cite{derby_spie_2026,subramanian_spie_2026}, all in Steward Observatory.  SCoOB is notable for including operations in a vacuum test chamber in preparation for spaceflight\cite{van_gorkom_spie_2026}. The XWCTk is also the baseline software for the Netherlands EXoplanet Testbed (NEXT) under development at Leiden University\cite{landman_spie_2026_2}.
       
\subsection{Spaceflight}

The XWCTk has been selected as the software to control the Extrasolar Coronagraph(ESC)\cite{miller_spie_2026} and the Widefield Context Camera (WCC)\cite{kautz_spie_2026} on the Lazuli space mission\cite{stefansson_spie_2026}.  These instruments will work together to manage acquisition and imaging with a distributed compute system.  Significant effort is underway to prepare for spaceflight.  This includes improving test coverage and increasing the support for automation.

\section{Conclusion}

We have presented the XWCTk, a software system designed to manage the complexity of ExAO and Coronagraph WFS\&C systems.  It has been in use on-sky as part of the MagAO-X instrument on the Magellan Clay telescope since 2019.  Since then it has been adopted by several testbeds, ground-based instrument projects, and is now being prepared for spaceflight applications.  Perhaps more importantly, it is routinely being used for ground-breaking discoveries\cite{2025ApJ...990L...9C} and experimental demonstrations\cite{2021JATIS...7b9001H}.  

\acknowledgments 

We are very grateful for support from the NSF MRI Award \#1625441. The Phase II upgrade program is made possible by the generous support of the Heising-Simons Foundation. MagAO-X uses the CACAO software package, which is supported by NSF Award \#2410616. Portions of this research were supported by generous anonymous philanthropic donations to the Steward Observatory of the College of Science at the University of Arizona.

\bibliography{report} 

\begin{thebibliography}{10}

\bibitem{2018ARA&A..56..315G}
{Guyon}, O., ``{Extreme Adaptive Optics},'' {\em ARA\&A}~{\bf 56},  315--355
  (Sept. 2018).

\bibitem{2014PASP..126..586S}
{Singh}, G., {Martinache}, F., {Baudoz}, P., {Guyon}, O., {Matsuo}, T.,
  {Jovanovic}, N., and {Clergeon}, C., ``{Lyot-based Low Order Wavefront Sensor
  for Phase-mask Coronagraphs: Principle, Simulations and Laboratory
  Experiments},'' {\em PASP}~{\bf 126},  586 (June 2014).

\bibitem{2024SPIE13097E..09M}
{Males}, J.~R., {Close}, L.~M., {Haffert}, S.~Y., {Kautz}, M.~Y., {Kueny}, J.,
  {Long}, J.~D., {McEwen}, E., {Swimmer}, N., {Bailey}, J.~I., {Foster}, W.,
  {Mazin}, B.~A., {Pearce}, L., {Liberman}, J., {Twitchell}, K., {Weinberger},
  A.~J., {Guyon}, O., {Hedglen}, A.~D., {McLeod}, A., {Roberts}, R., {Van
  Gorkom}, K., {Li}, J., {Doty}, I., and {Gasho}, V., ``{MagAO-X: commissioning
  results and status of ongoing upgrades},'' in [{\em Adaptive Optics Systems
  IX}{\nolinebreak\hspace{0.1em}]},  {Jackson}, K.~J., {Schmidt}, D., and
  {Vernet}, E., eds., {\em Society of Photo-Optical Instrumentation Engineers
  (SPIE) Conference Series} {\bf 13097},  1309709 (Aug. 2024).

\bibitem{lucas_spie_2026}
{Lucas}, M. et~al., ``{Visible-light high-contrast polarimetric imaging with
  MagAO-X: characterization and early results},'' in [{\em Ground-based and
  Airborne Instrumentation for Astronomy XI}{\nolinebreak\hspace{0.1em}]},
  {\em Proc. SPIE},  to appear (2026).

\bibitem{tonucci_spie_2026_1}
{Tonucci}, E. et~al., ``{First on-sky focal plane wavefront control
  demonstration with the self-coherent camera},'' in [{\em Adaptive Optics
  X}{\nolinebreak\hspace{0.1em}]},  {\em Proc. SPIE},  to appear (2026).

\bibitem{close_spie_2026_1}
{Close}, L.~M. et~al., ``{A review of high-contrast detection of accreting
  proto-planets with visible ExAO at H-alpha: the MagAO-X MaXProtoPlanetS
  survey results},'' in [{\em Adaptive Optics X}{\nolinebreak\hspace{0.1em}]},
  {\em Proc. SPIE},  to appear (2026).

\bibitem{twitchell_spie_2026}
{Twitchell}, K. et~al., ``{Closed-loop atmospheric dispersion correction for
  high contrast imaging with MagAO-X},'' in [{\em Adaptive Optics
  X}{\nolinebreak\hspace{0.1em}]},  {\em Proc. SPIE},  to appear (2026).

\bibitem{mars_spie_2026_1}
{Mars}, M. et~al., ``{Physics-informed digital twin for forward modelling
  low-order aberrations in high-contrast imaging},'' in [{\em Adaptive Optics
  X}{\nolinebreak\hspace{0.1em}]},  {\em Proc. SPIE},  to appear (2026).

\bibitem{mcewen_spie_2026_1}
{McEwen}, E. et~al., ``{Comparing real-time optical gain measurement and
  methods on MagAO-X},'' in [{\em Adaptive Optics
  X}{\nolinebreak\hspace{0.1em}]},  {\em Proc. SPIE},  to appear (2026).

\bibitem{mcewen_spie_2026_2}
{McEwen}, E. et~al., ``{$\beta$-pic b in the blue: high contrast detections
  from 0.6-0.9$\mu$m with MagAO-X},'' in [{\em Adaptive Optics
  X}{\nolinebreak\hspace{0.1em}]},  {\em Proc. SPIE},  to appear (2026).

\bibitem{mars_spie_2026_2}
{Mars}, M. et~al., ``{Non-linear low-order wavefront control with a
  physics-based digital twin: on-sky results from MagAO-X},'' in [{\em Adaptive
  Optics X}{\nolinebreak\hspace{0.1em}]},  {\em Proc. SPIE},  to appear (2026).

\bibitem{kueny_spie_2026_1}
{Kueny}, J. et~al., ``{WindsoCC: reconstructing the wind-driven halo in MagAO-X
  images using wavefront sensor telemetry},'' in [{\em Adaptive Optics
  X}{\nolinebreak\hspace{0.1em}]},  {\em Proc. SPIE},  to appear (2026).

\bibitem{landman_spie_2026_1}
{Landman}, R. et~al., ``{No need to modulate: on-sky results of a neural
  network enhanced pyramid wavefront sensor and prospects for the ELT’s},''
  in [{\em Adaptive Optics X}{\nolinebreak\hspace{0.1em}]},  {\em Proc. SPIE},
  to appear (2026).

\bibitem{liberman_spie_2026}
{Liberman}, J. et~al., ``{Freezing the speckles: focal plane wavefront sensing
  with the Spatially Clipped Self-Coherent Camera},'' in [{\em Adaptive Optics
  X}{\nolinebreak\hspace{0.1em}]},  {\em Proc. SPIE},  to appear (2026).

\bibitem{kueny_spie_2026}
{Kueny}, J. et~al., ``{ffortissimo: a novel JAX-based pipeline for freeform
  forward modeling of circumstellar disks},'' in [{\em Adaptive Optics
  X}{\nolinebreak\hspace{0.1em}]},  {\em Proc. SPIE},  to appear (2026).

\bibitem{long_spie_2026}
{Long}, J. et~al., ``{What the speck is that? Improving exoplanet imaging
  sensitivity by combining machine learning and physical models},'' in [{\em
  Adaptive Optics X}{\nolinebreak\hspace{0.1em}]},  {\em Proc. SPIE},  to
  appear (2026).

\bibitem{haffert_spie_2026_1}
{Haffert}, S. et~al., ``{On-sky demonstration of self-learning predictive
  control with MagAO-X},'' in [{\em Adaptive Optics
  X}{\nolinebreak\hspace{0.1em}]},  {\em Proc. SPIE},  to appear (2026).

\bibitem{gonzalez_spie_2026}
{Gonzalez}, M. et~al., ``{Visible light spectroscopy with extreme adaptive
  optics: a pipeline for the low-resolution mode of VIS-X},'' in [{\em Adaptive
  Optics X}{\nolinebreak\hspace{0.1em}]},  {\em Proc. SPIE},  to appear (2026).

\bibitem{patel_spie_2026}
{Patel}, D. et~al., ``{Demonstration of simultaneous PIAA-coronagraphy and
  wavefront sensing using a single metasurface-based focal-plane optic},'' in
  [{\em Advances in Optical and Mechanical Technologies for Telescopes and
  Instrumentation VII}{\nolinebreak\hspace{0.1em}]},  {\em Proc. SPIE},  to
  appear (2026).

\bibitem{tonucci_spie_2026_2}
{Tonucci}, E. et~al., ``{Sub-diffraction-limited coronagraphic imaging with
  nano-printed PIAACMC phase masks},'' in [{\em Advances in Optical and
  Mechanical Technologies for Telescopes and Instrumentation
  VII}{\nolinebreak\hspace{0.1em}]},  {\em Proc. SPIE},  to appear (2026).

\bibitem{johnson_spie_2026}
{Johnson}, P. et~al., ``{Sensor fusion on MagAO-X: real time vibration control
  using accelerometers},'' in [{\em Adaptive Optics
  X}{\nolinebreak\hspace{0.1em}]},  {\em Proc. SPIE},  to appear (2026).

\bibitem{hom_spie_2026}
{Hom}, J. et~al., ``{The Roman Coronagraph Community Participation Program:
  pre-launch reference star list and impact of reference star properties on
  post-processing performance},'' in [{\em Space Telescopes and Instrumentation
  2026: Optical, Infrared, and Millimeter Wave}{\nolinebreak\hspace{0.1em}]},
  {\em Proc. SPIE},  to appear (2026).

\bibitem{indi_specs}
{Downey}, E.~C., ``{INDI: Instrument-Neutral Distributed Interface},'' tech.
  rep. (6 2007).

\bibitem{2018SPIE10703E..1EG}
{Guyon}, O., {Sevin}, A., {Gratadour}, D., {Bernard}, J., {Ltaief}, H.,
  {Sukkari}, D., {Cetre}, S., {Skaf}, N., {Lozi}, J., {Martinache}, F.,
  {Clergeon}, C., {Norris}, B., {Wong}, A., and {Males}, J., ``{The compute and
  control for adaptive optics (CACAO) real-time control software package},'' in
  [{\em Adaptive Optics Systems VI}{\nolinebreak\hspace{0.1em}]},  {Close},
  L.~M., {Schreiber}, L., and {Schmidt}, D., eds., {\em Society of
  Photo-Optical Instrumentation Engineers (SPIE) Conference Series} {\bf
  10703},  107031E (July 2018).

\bibitem{deo_spie_2026}
{Deo}, V. et~al., ``{CACAO++: scheduling user experience as the core of
  adaptive optics real-time computers},'' in [{\em Adaptive Optics Systems
  X}{\nolinebreak\hspace{0.1em}]},  {\em Proc. SPIE},  to appear (2026).

\bibitem{2018SPIE10703E..42P}
{Por}, E.~H., {Haffert}, S.~Y., {Radhakrishnan}, V.~M., {Doelman}, D.~S., {van
  Kooten}, M., and {Bos}, S.~P., ``{High Contrast Imaging for Python (HCIPy):
  an open-source adaptive optics and coronagraph simulator},'' in [{\em
  Adaptive Optics Systems VI}{\nolinebreak\hspace{0.1em}]},  {Close}, L.~M.,
  {Schreiber}, L., and {Schmidt}, D., eds., {\em Society of Photo-Optical
  Instrumentation Engineers (SPIE) Conference Series} {\bf 10703},  1070342
  (July 2018).

\bibitem{2022SPIE12185E..3PL}
{Long}, J.~D., {Males}, J.~R., {Haffert}, S.~Y., {Close}, L.~M., {Morzinski},
  K.~M., {Van Gorkom}, K., {Lumbres}, J., {Foster}, W., {Hedglen}, A., {Kautz},
  M., {Rodack}, A., {Schatz}, L., {Miller}, K., {Doelman}, D., {Bos}, S.~P.,
  {Kenworthy}, M.~A., {Snik}, F., and {Otten}, G. P.~P.~L., ``{XPipeline:
  starlight subtraction at scale for MagAO-X},'' in [{\em Adaptive Optics
  Systems VIII}{\nolinebreak\hspace{0.1em}]},  {Schreiber}, L., {Schmidt}, D.,
  and {Vernet}, E., eds., {\em Society of Photo-Optical Instrumentation
  Engineers (SPIE) Conference Series} {\bf 12185},  121853P (Aug. 2022).

\bibitem{2025A&A...696L...1L}
{Landman}, R., {Haffert}, S.~Y., {Long}, J.~D., {Males}, J.~R., {Close}, L.~M.,
  {Foster}, W.~B., {Van Gorkom}, K., {Guyon}, O., {Hedglen}, A.~D., {Johnson},
  P.~T., {Kautz}, M.~Y., {Kueny}, J.~K., {Li}, J., {Liberman}, J., {Lumbres},
  J., {McEwen}, E.~A., {McLeod}, A., {Schatz}, L., {Tonucci}, E., and
  {Twitchell}, K., ``{Making the unmodulated pyramid wavefront sensor smart:
  II. First on-sky demonstration of extreme adaptive optics with deep
  learning},'' {\em A\&A}~{\bf 696},  L1 (Apr. 2025).

\bibitem{montoya_spie_2026}
{Montoya}, M.~M. et~al., ``{The MMT adaptive optics exoplanet characterization
  system (MAPS) update},'' in [{\em Adaptive Optics Systems
  X}{\nolinebreak\hspace{0.1em}]},  {\em Proc. SPIE},  to appear (2026).

\bibitem{males_spie_2026_1}
{Males}, J.~R. et~al., ``{The final design of GMagAO-X: high-contrast imaging
  at first-light of the GMT},'' in [{\em Ground-based and Airborne
  Instrumentation for Astronomy XI}{\nolinebreak\hspace{0.1em}]},  {\em Proc.
  SPIE},  to appear (2026).

\bibitem{close_spie_2026_2}
{Close}, L.~M. et~al., ``{The GMagAO-X facility ExAO coronagraphic instrument
  for the Giant Magellan Telescope: the first FDR level optomechanical design
  for an ELT ExAO system},'' in [{\em Adaptive Optics
  X}{\nolinebreak\hspace{0.1em}]},  {\em Proc. SPIE},  to appear (2026).

\bibitem{haffert_spie_2026_2}
{Haffert}, S. et~al., ``{The final design of GMagAO-X: the wavefront sensing
  and control (WFS\&C) architecture of GMagAO-X},'' in [{\em Adaptive Optics
  X}{\nolinebreak\hspace{0.1em}]},  {\em Proc. SPIE},  to appear (2026).

\bibitem{giorgetti_spie_2026}
{Giorgetti}, T. et~al., ``{A compact atmospheric dispersion corrector mechanism
  for GMT’s GMagAO-X exoplanet instrument},'' in [{\em Advances in Optical
  and Mechanical Technologies for Telescopes and Instrumentation
  VII}{\nolinebreak\hspace{0.1em}]},  {\em Proc. SPIE},  to appear (2026).

\bibitem{2022JATIS...8d9001S}
{Schatz}, L., {Codona}, J., {Long}, J.~D., {Males}, J.~R., {Pullen}, W.,
  {Lumbres}, J., {Van Gorkom}, K., {Chambouleyron}, V., {Close}, L.~M.,
  {Correia}, C., {Fauvarque}, O., {Fusco}, T., {Guyon}, O., {Hart}, M.,
  {Janin-Potiron}, P., {Johnson}, R., {Jovanovic}, N., {Mateen}, M., {Sauvage},
  J.-F., and {Neichel}, B., ``{Three-sided pyramid wavefront sensor, part II:
  preliminary demonstration on the new comprehensive adaptive optics and
  coronagraph test instrument testbed},'' {\em Journal of Astronomical
  Telescopes, Instruments, and Systems}~{\bf 8},  049001 (Oct. 2022).

\bibitem{blomquist_spie_2026}
{Blomquist}, S. et~al., ``{Tiny Observatory for Telescope Optimization (TOTO):
  testing algorithms for autonomous on-orbit alignment for space-based
  telescope systems},'' in [{\em Space Telescopes and Instrumentation 2026:
  Optical, Infrared, and Millimeter Wave}{\nolinebreak\hspace{0.1em}]},  {\em
  Proc. SPIE},  to appear (2026).

\bibitem{derby_spie_2026}
{Derby}, K. et~al., ``{The Space Coronagraph Optical Bench (SCoOB): 9.
  improvements to adjoint electric field conjugation and a demonstration in
  broadband light},'' in [{\em Space Telescopes and Instrumentation 2026:
  Optical, Infrared, and Millimeter Wave}{\nolinebreak\hspace{0.1em}]},  {\em
  Proc. SPIE},  to appear (2026).

\bibitem{subramanian_spie_2026}
{Subramanian}, S.~K. et~al., ``{The Space Coronagraph Optical Bench (SCoOB):
  10. dark zone maintenance for space-based high contrast imaging testbed},''
  in [{\em Space Telescopes and Instrumentation 2026: Optical, Infrared, and
  Millimeter Wave}{\nolinebreak\hspace{0.1em}]},  {\em Proc. SPIE},  to appear
  (2026).

\bibitem{van_gorkom_spie_2026}
{Van Gorkom}, K. et~al., ``{The Space Coronagraph Optical Bench (SCoOB): 11.
  status of ongoing upgrades and vacuum performance},'' in [{\em Space
  Telescopes and Instrumentation 2026: Optical, Infrared, and Millimeter
  Wave}{\nolinebreak\hspace{0.1em}]},  {\em Proc. SPIE},  to appear (2026).

\bibitem{landman_spie_2026_2}
{Landman}, R. et~al., ``{The Netherlands EXoplanet Testbed (NEXT): goals and
  opto-mechanical design},'' in [{\em Adaptive Optics
  X}{\nolinebreak\hspace{0.1em}]},  {\em Proc. SPIE},  to appear (2026).

\bibitem{miller_spie_2026}
{Miller}, K. et~al., ``{The Lazuli Space Observatory’s ExtraSolar
  Coronagraph},'' in [{\em Space Telescopes and Instrumentation 2026: Optical,
  Infrared, and Millimeter Wave}{\nolinebreak\hspace{0.1em}]},  {\em Proc.
  SPIE},  to appear (2026).

\bibitem{kautz_spie_2026}
{Kautz}, M. et~al., ``{Technical status of the widefield context camera in the
  Lazuli Space Observatory},'' in [{\em Space Telescopes and Instrumentation
  2026: Optical, Infrared, and Millimeter Wave}{\nolinebreak\hspace{0.1em}]},
  {\em Proc. SPIE},  to appear (2026).

\bibitem{stefansson_spie_2026}
{Stefansson}, G.~K. et~al., ``{The Lazuli Space Observatory},'' in [{\em Space
  Telescopes and Instrumentation 2026: Optical, Infrared, and Millimeter
  Wave}{\nolinebreak\hspace{0.1em}]},  {\em Proc. SPIE},  to appear (2026).

\bibitem{2025ApJ...990L...9C}
{Close}, L.~M., {van Capelleveen}, R.~F., {Weible}, G., {Wagner}, K.,
  {Haffert}, S.~Y., {Males}, J.~R., {Ilyin}, I., {Kenworthy}, M.~A., {Li}, J.,
  {Long}, J.~D., {Ertel}, S., {Ginski}, C., {Weinberger}, A.~J., {Follette},
  K., {Liberman}, J., {Twitchell}, K., {Johnson}, P., {Kueny}, J., {Apai}, D.,
  {Doyon}, R., {Foster}, W., {Gasho}, V., {Van Gorkom}, K., {Guyon}, O.,
  {Kautz}, M.~Y., {McLeod}, A., {McEwen}, E., {Pearce}, L., {Schatz}, L.,
  {Hedglen}, A.~D., {Wu}, Y.-L., {Isbell}, J., {Power}, J., {Carlson}, J.,
  {Close}, E., {Tonucci}, E., and {Mars}, M., ``{Wide Separation Planets in
  Time (WISPIT): Discovery of a Gap H{\ensuremath{\alpha}} Protoplanet WISPIT
  2b with MagAO-X},'' {\em ApJL}~{\bf 990},  L9 (Sept. 2025).

\bibitem{2021JATIS...7b9001H}
{Haffert}, S.~Y., {Males}, J.~R., {Close}, L.~M., {Van Gorkom}, K., {Long},
  J.~D., {Hedglen}, A.~D., {Guyon}, O., {Schatz}, L., {Kautz}, M., {Lumbres},
  J., {Rodack}, A., {Knight}, J.~M., {Sun}, H., and {Fogarty}, K.,
  ``{Data-driven subspace predictive control of adaptive optics for
  high-contrast imaging},'' {\em Journal of Astronomical Telescopes,
  Instruments, and Systems}~{\bf 7},  029001 (Apr. 2021).

\end{thebibliography}
\bibliographystyle{spiebib} 

\end{document}